\documentclass{aa}
\usepackage{graphicx}
\usepackage{natbib}
\usepackage{amsmath}
\bibpunct{(}{)}{;}{a}{}{,}

\newcommand{\ffh}{\hat f}
\newcommand{\gh}{\hat g}
\newcommand{\Kh}{\hat K}
\newcommand{\zetah}{\hat{\zeta}}
\newcommand{\varthetah}{\hat{\vartheta}}
\newcommand{\Kbh}{\boldsymbol {\hat{K}}}
\newcommand{\fbh}{\boldsymbol {\hat{f}}}
\newcommand{\gbh}{\boldsymbol {\hat{g}}}
\newcommand{\oh}{\hat o}
\newcommand{\wh}{\hat w}

\newcommand{\fb}{\boldsymbol{f}}
\newcommand{\hb}{\boldsymbol{h}}
\newcommand{\gb}{\boldsymbol{g}}
\newcommand{\ob}{\boldsymbol{o}}
\newcommand{\wb}{\boldsymbol{w}}

\newcommand{\varrhob}{\boldsymbol{\varrho}}

\newcommand{\Ab}{\boldsymbol{A}}
\newcommand{\Db}{\boldsymbol{D}}
\newcommand{\Deltab}{\boldsymbol{\Delta}}
\newcommand{\Lambdab}{\boldsymbol{\Lambda}}
\newcommand{\amatb}{\boldsymbol {\cal A}}

\newcommand{\kmath}{\hat {\cal K}}

\begin{document}

\title{A simple but efficient algorithm \\ for multiple-image deblurring}

   \author{R. Vio\inst{1}
          \and
          J. Nagy\inst{2}
          \and
          L. Tenorio\inst{3}
          \and
          W. Wamsteker\inst{4}
          }

   \offprints{R. Vio}

   \institute{Chip Computers Consulting s.r.l., Viale Don L.~Sturzo 82,
              S.Liberale di Marcon, 30020 Venice, Italy\\
              ESA-VILSPA, Apartado 50727, 28080 Madrid, Spain\\
              \email{robertovio@tin.it}
         \and
                  Department of Mathematics and Computer Science, Emory University,
                        Atlanta, GA 30322, USA. \\
              \email{nagy@mathcs.emory.edu}
         \and
              Department of Mathematical and Computer Sciences, Colorado School
                  of Mines, Golden CO 80401, USA \\
              \email{ltenorio@Mines.EDU}
        \and
             ESA-VILSPA, Apartado 50727, 28080 Madrid, Spain\\
             \email{willem.wamsteker@esa.int}
             }

\date{Received .............; accepted ................}

\abstract{We consider the simultaneous deblurring of a set of noisy images whose point spread functions
are different but known and spatially invariant, and the noise is Gaussian. Currently available iterative algorithms
that are typically used for this type of problem are computationally expensive, which makes
their application for very large images impractical.
We present a simple extension of a classical least-squares (LS) method where
the multi-image deblurring is efficiently reduced to a computationally efficient single-image deblurring.
In particular, we show that it is possible to remarkably improve the ill-conditioning of the LS problem
by means of stable operations on the corresponding normal equations, which in turn speed up
the convergence rate of the iterative algorithms.
The performance  and limitations of the method are analyzed through
numerical simulations. Its connection with a column weighted least-squares approach
is also considered in an appendix.
}
\titlerunning{Multiple image deblurring}
\authorrunning{R. Vio, J. Nagy, L. Tenorio, \& W. Wamsteker}
\maketitle

\section{Introduction}

An important problem in image processing is the
simultaneous deblurring of a set of observed images (of a fixed object),
where each image is degraded by a different spatially
invariant point spread function (PSF).
For example, data obtained
by the Large Binocular Telescope (LBT). This instrument consists of two $8.4 {\rm m}$ mirrors on a common mount
with a spacing of $14.4 {\rm m}$ between the centers \citep{ang98}. For a given orientation of the telescope, the
diffraction-limited resolution
along the center-to-center baseline is equivalent to that of a $22.8 {\rm m}$ mirror, while the resolution along
the perpendicular
direction is that of a single $8 {\rm m}$ mirror. A possible way to obtain an image with an improved and uniform spatial
resolution is to simultaneously deconvolve the images taken with different orientations of the telescope.
Another example is the multi-frequency observations of the Cosmic Microwave Background (CMB) obtained from
satellites such as {\it PLANCK}.
In fact, although some other physical components are present at the microwave frequencies, the images taken at high
galactic latitudes are expected to be almost entirely  dominated by the CMB contribution in a large range of observing
frequencies \citep[e.g., see][]{sto02}. Since the PSFs corresponding to the images obtained at the various
frequencies can be quite dissimilar, simultaneous deblurring can be useful for improving the extraction
of CMB information from the data.

Although a vast literature on the classical problem of single-image deblurring is available, much less has been
published on multiple-image deblurring. An excellent general review is
\citet{ber00b}.  In addition, \citet{teg99} provides some material more directly related to CMB studies.
But one of the more serious
limitations of the methods currently available is that, although able to provide satisfactory results, they are
quite expensive in terms of computational requirements. Thus, these methods may
be quite difficult to use for the very large images ($\approx 10^6 - 10^8$ pixels) expected from LBT and {\it PLANCK}
observations.
It is clear that more efficient methods must be developed; the aim of this paper is to provide one such approach.

In Sect.~\ref{sec:formalization} we formalize the problem and propose an efficient solution
in Sect.~\ref{sec:efficient}. Some of its possible limitations are considered in
Sect.~\ref{sec:problems}. In Sect.~\ref{sec:experiments} we study its performance through numerical experiments,
and also compare it to that of other methods currently available in the literature.
Finally, Sect.~\ref{sec:conclusions} closes with some final comments and conclusions.

\section{Formalization of the problem} \label{sec:formalization}

Suppose one has $p$ observed images, $g_j(x,y)$, $j=1,2, \ldots, p$, of a fixed two-dimensional object $f_0(x,y)$,
each of which is degraded by a different spatially invariant blurring operator.  That is,
\begin{equation} \label{eq:model1}
g_j(x,y) = \iint K_j (x-x', y-y') f_0(x',y') dx' dy',
\end{equation}
where, for each $j$, $K_j(x,y)$ is a space invariant PSF.

The image restoration problem of interest is to find an estimate $f(x,y)$ of $f_0(x,y)$ given
the observed images $\{ g_j(x,y) \}$ and the known PSFs $\{ K_j(x,y) \}$.

In practice, the experimental images are discrete $N \times N$ arrays of pixels
(for simplicity, we assume square images) contaminated by noise. Therefore model~(\ref{eq:model1})
has to be recast in the form
\begin{equation} \label{eq:model2}
g_j(m,n) = \sum_{m',n'=0}^{N-1} K_j(m-m', n-n') f_0(m',n') + w_j(m,n),
\end{equation}
where $w_j(m,n)$ is additive Gaussian white noise. For the moment, we assume constant standard deviations
across images: $\sigma_{w_j} = \sigma_w$.

If the central peak (i.e., support) of each PSF is much smaller than the image, and
if the object does not have structures near the image boundaries, then the convolution
product in Eq.~(\ref{eq:model2}) can be well approximated by cyclic convolution. As is well known,
such an approximation is quite useful since it permits rewriting Eq.~({\ref{eq:model2}) as
\begin{equation} \label{eq:model}
\gh_j(m,n) = \Kh_j(m,n) \ffh_0(m,n) + \wh_j(m,n),
\end{equation}
where the symbol ``$~\hat{}~$'' denotes the Discrete Fourier Transform (DFT).

One approach to compute an estimate of $f_0$ is to pose the image restoration
problem as a least-squares (LS) problem, which
is equivalent to a maximum likelihood approach in the case of white
Gaussian noise.  In order to outline the LS approach, it
is helpful to use matrix-vector notation:
Eq.~(\ref{eq:model2}) can be rewritten as
\begin{equation} \label{eq:modela}
\gb_j = \Ab_j \fb_0 + \wb_j,
\end{equation}
where the arrays $\gb$, $\fb_0$ and $\wb_j$ are obtained by column stacking $g_j(m,n)$,
$f_0(m,n)$, and $w_j(m,n)$, respectively. $\Ab_j$ is the block-circulant matrix defined by cyclic convolution
with the $j$th PSF. The LS estimate $\fb$ can then be expressed in the
form:
\begin{equation} \label{eq:min2}
\fb = \arg\min \,\sum_{j=1}^{p} \Vert \Ab_j \fb - \gb_j \Vert^2 .
\end{equation}
It is not difficult to see that $\fb$ is the solution of the {normal equations}
\begin{equation} \label{eq:model3}
\sum_{j=1}^p \Ab_j^T \Ab_j \fb = \sum_{j=1}^p \Ab_j^T \gb_j,
\end{equation}
where $\Ab_j^T$ denotes the transpose of the block-circulant matrix $\Ab_j$.
The DFT of both sides of Eq.~(\ref{eq:model3}) provides
\begin{equation} \label{eq:single}
\ffh(m,n) \sum_{j=1}^p | \Kh_j(m,n) |^2 = \sum_{j=1}^p \Kh_j^*(m,n) \gh_j(m,n) = \varthetah(m,n),
\end{equation}
with symbol ``$~{}^*~$'' denoting complex conjugation. This equation shows that the multi-image
LS deblurring problem (\ref{eq:min2}) is equivalent to classical one-image LS deblurring
where the ``observed'' image is $\gh(m,n) = \varthetah(m,n)$ and the PSF is
$\Kh(m,n) = \sum_{j=1}^p | \Kh_j(m,n) |^2$.
As is well known, the direct LS estimate
given by the inverse DFT of
\begin{equation} \label{eq:solution1}
\ffh(m,n) = \frac{\sum_j \Kh_j^*(m,n) \gh_j(m,n)}{\sum_j | \Kh_j(m,n) |^2}
\end{equation}
can be very unstable. For white noise, the mean square error (MSE) of (\ref{eq:solution1}) is
\begin{equation} \label{eq:variance1}
{\rm E} [ \Vert \fb - \fb_0 \Vert^2 ] =  \sigma^2_w \sum_{m,n} \frac{1}{\sum_j | \Kh_j(m,n) |^2}.
\end{equation}
The quantities $ \Kh_j(m,n) $ can be very close to zero at high frequencies because the PSFs
are almost band-limited in practice. Consequently, the MSE can be very large, which indicates that the LS estimate
is unstable. The main task of any deblurring method consists
of stabilizing the solution by somehow limiting the values of such terms.
Various methods have been proposed
in the literature \citep[e.g., see the review by ][]{ber00b}. For example, the Tikhonov estimate $\fb^{(\mu)}$ is
defined as
\begin{equation} \label{eq:min3}
\fb^{(\mu)} = \arg\min \,\left[\,\sum_{j=1}^{p} \Vert \Ab_j \fb - \gb_j \Vert^2  + \mu \Vert \fb \Vert^2\,\right] ,
\end{equation}
where $\mu$ is a positive scalar called the {regularization parameter}. The additional penalty term (weighted by $\mu$)
has the effect of limiting the energy of $\fb$.
In fact, it is possible to see that
\begin{equation}
\ffh^{(\mu)}(m,n) = \frac{\sum_j \Kh_j^*(m,n) \gh_j(m,n)}{\mu + \sum_j | \Kh_j(m,n) |^2},
\end{equation}
where it is evident that the parameter $\mu$ smooths out the influence of the smallest values of $\Kh_j(m,n)$. If
both the numerator and the
second term in the denominator are computed and stored, then for each value of $\mu$ the computation
of $\fb^{(\mu)}$ requires $N^2$ divisions + $N^2$ sums + one two-dimensional DFT.
This means that the Tikhonov method is computationally efficient, but it is not always flexible enough
to allow the implementation of constraints such as positivity of the solution.

Iterative methods provide an alternative methodology that
exploits the semiconvergence property of iterative inversion algorithms and can be easily implemented
with many different types of constraints on the solution. Two examples (which provide constrained positive solutions)
are the Projected Landweber (PL) method and the Iterative Space Reconstruction Algorithm (ISRA) \citep[see ][]{ber00b}.
But, although flexible, these methods are expensive
from the computational point of view, especially in situations of slow convergence (for each iteration,
a couple of two-dimensional DFTs is required).

\section{A more efficient approach} \label{sec:efficient}

The main limitation of the methods mentioned in the previous section is that they are based on the
normal equations (\ref{eq:single}). The resulting deblurring problem is much more ill-conditioned
(i.e., the PSF is much broader)  than implied by each of the
models~(\ref{eq:model2}) because the coefficients $\Kh(m,n)$ in Eq.~(\ref{eq:single}) are squared.
The most important consequence is that the convergence of the iterative algorithms can be remarkably
slowed with obvious consequence on the computational cost. However, by multiplying both sides of
Eq.~(\ref{eq:single}) by a function $\Kh_{j_0}(m,n)$, we can rewrite Eq.~(\ref{eq:single}) in the form
\begin{equation} \label{eq:solution2}
\begin{aligned}
& \Kh_{j_0}(m,n) \ffh(m,n) \\
& = \frac{\gh_{j_0}(m,n) + \sum_{j \neq j_0} [ \Kh_j^*(m,n) /  \Kh_{j_0}^*(m,n)] \gh_j(m,n)}
{1 + \sum_{j \neq j_0} [ | \Kh_j(m,n) |^2 / | \Kh_{j_0}(m,n) |^2]} \\
& = \zetah(m,n),
\end{aligned}
\end{equation}
where
\begin{equation} \label{eq:solution2k}
\Kh_{j_0}(m,n) = \max [ \Kh_1(m,n), \Kh_2(m,n), \ldots, \Kh_p(m,n) ].
\end{equation}
Here the operator $\max[~]$ extracts the element in the array with the largest modulus, and $j_0$ is the value of
the corresponding index $j$. In the case that two or more PSFs whose DFTs, for a given frequency $(m,n)$, have the same
modulus, it is sufficient to choose one of the $\Kh_j(m,n)$ according to a rule that preserves the symmetry
of the DFT of a real PSF.  As already noted for model~(\ref{eq:single}), model~(\ref{eq:solution2}) is also
equivalent to a classical LS deblurring, but now  the ``observed'' image is $\zetah(m,n)$ and the PSF is
$\Kh_{j_0}(m,n)$.

By construction, $\Kh_{j_0}(m,n)$ is the PSF whose energy
distribution is maximally spread in the Fourier domain and therefore it corresponds to a PSF
whose energy distribution is maximally concentrated in the spatial domain.
\begin{figure}
        \resizebox{\hsize}{!}{\includegraphics{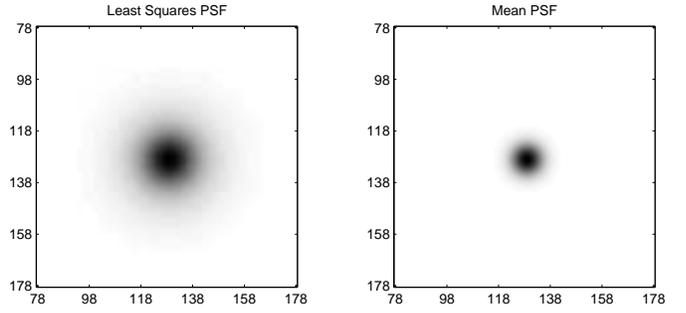}}
        \caption{Comparison between the PSF corresponding to Eq.~(\ref{eq:single}) (left panel) with that
	  corresponding to Eq.~(\ref{eq:solution2}) (right panel) for the numerical experiments presented in
	  Sect.~\ref{sec:experiments}.}
        \label{fig:psfs}
\end{figure}
\begin{figure}
        \resizebox{\hsize}{!}{\includegraphics{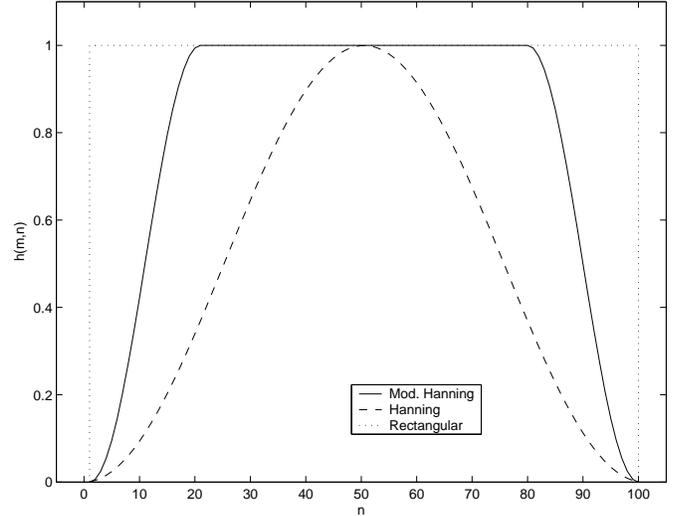}}
        \caption{Slice comparison of the two-dimensional classical Hanning and rectangular windows
        with the modified Hanning window ($N_w=20$).}
        \label{fig:windows}
\end{figure}
This means that $\zetah(m,n)$ can be considered a sort of deblurred version of $\varthetah(m,n)$
(see Fig.~\ref{fig:psfs}).
In other words, the conditioning of system~(\ref{eq:solution2}) is much better than that of
system~(\ref{eq:single}) (this is because $\Kh_{j_0}(m,n)$ is not obtained by squaring the coefficients $\Kh_j(m,n)$).
The important point is that this result has been obtained through stable operations. From these considerations,
improvements in the computational costs are to be expected.
In addition, by formulating the problem in terms of single-image deblurring, there is a gain
in flexibility as standard deblurring algorithms can be used.

In Appendix \ref{sec:ls}, it is shown that
$\zetah(m,n)$ is equivalently obtained by transforming the LS problem~(\ref{eq:min2})
into a column weighted LS problem.

\section{Two potential problems} \label{sec:problems}

Although simple and potentially very effective, the procedure presented in the previous sections has two
potential drawbacks.

The first concerns the statistical properties of the noise component of
$\zeta(m,n)$. In particular, even if the original noise $\wb_j$ is white (i.e., a process with flat spectrum),
the noise component of $\zeta(m,n)$ has a power-spectrum proportional to
$|\Kh_{j_0}(m,n) |^2/\sum_j | \Kh_j(m,n) |^2$.
This is a direct consequence of having used the normal equations~(\ref{eq:single}) to solve the LS
problem~(\ref{eq:min2}) and may have deleterious effects on automatic methods for stopping
the iterations or estimating the regularization parameters since they assume that noise is white.
For example, in our simulations the generalized cross validation (GCV)
method \citep[which may be used to estimate the optimal value of $\mu$ in the
Tikhonov approach, see][]{vog02} consistently produced values that were too small.

\begin{figure}
        \resizebox{\hsize}{!}{\includegraphics{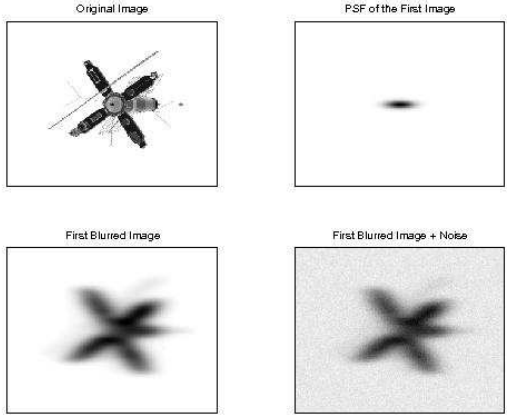}}
        \caption{Original image, blurred version and
        blurred + noise version of the first image of the set (see text) and corresponding PSF. The
        images are $256 \times 256$ pixels. }
        \label{fig:sat_iterc1_50}
\end{figure}
\begin{figure}
        \resizebox{\hsize}{!}{\includegraphics{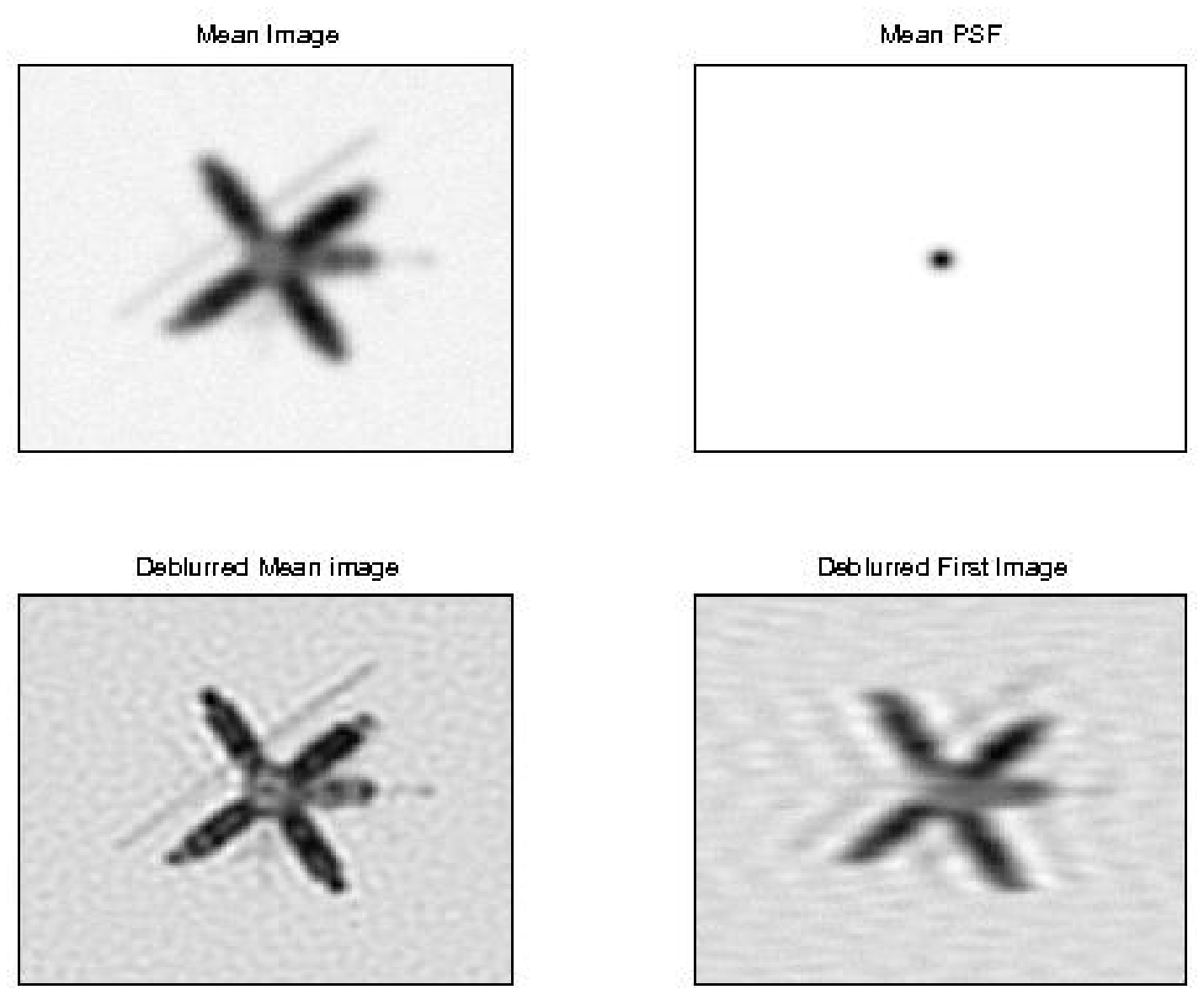}}
        \caption{Mean image $\zeta(m,n)$ and corresponding PSF for the image shown in
	  Fig.~\ref{fig:sat_iterc1_50}, deblurred version of $\zeta(m,n)$,
        and deblurred version of the first image of the set (see text). The
        deblurring was done with CGLS. The images shown are the ones with the smallest
	  standard deviation of the true residuals; for the mean image and first image of the set
        these are, respectively,
        $6.57 \times 10^{-2}$ and $8.67 \times 10^{-2}$.}
        \label{fig:sat_iterc2_50}
\end{figure}
\begin{figure}
        \resizebox{\hsize}{!}{\includegraphics{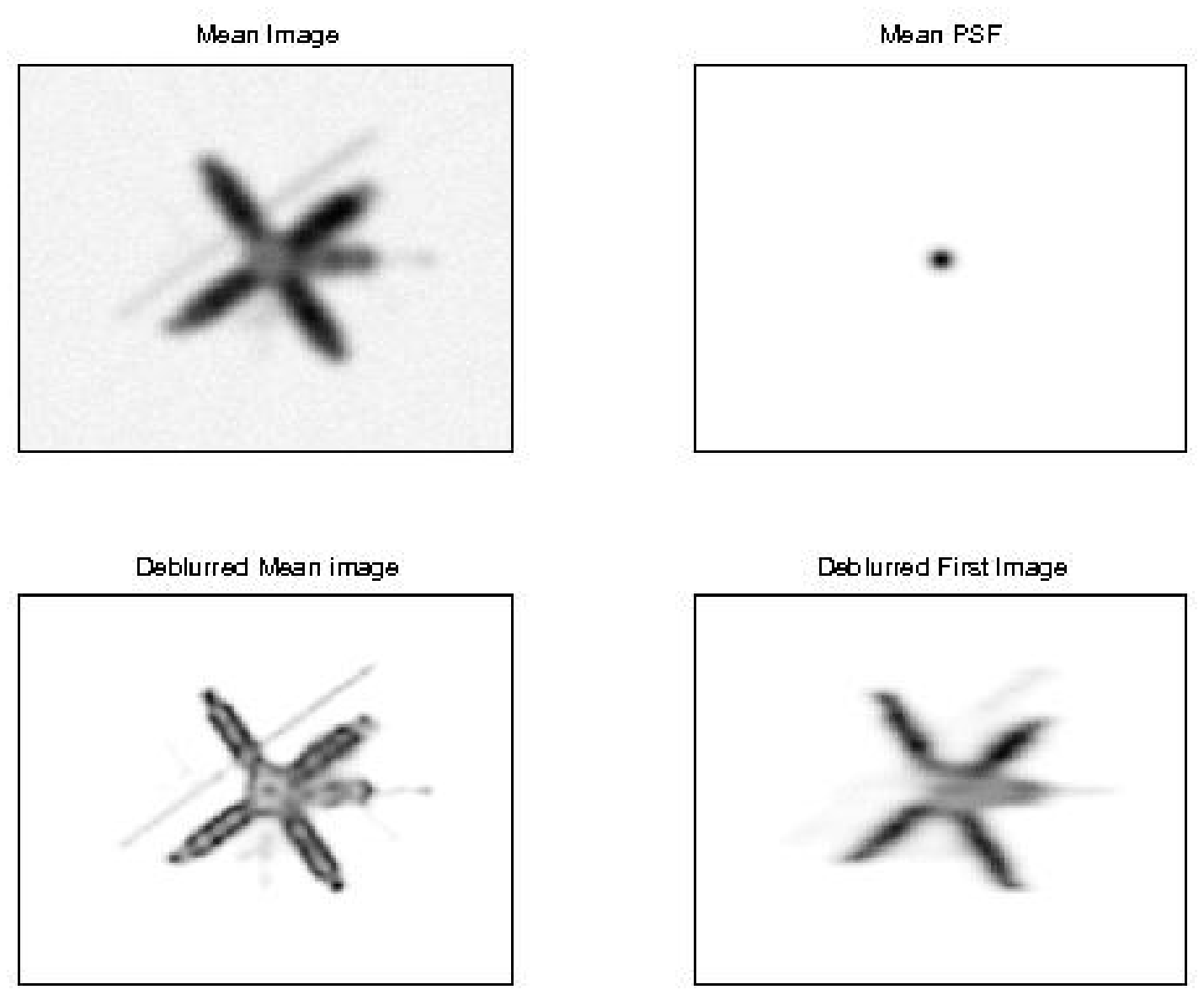}}
        \caption{Mean image $\zeta(m,n)$ and corresponding PSF for the image shown in
	  Fig.~\ref{fig:sat_iterc1_50}, deblurred version of $\zeta(m,n)$,
        and deblurred version of the first image of the set (see text). The
        deblurring was done with MRNSD. The images shown are the ones with the smallest
	  standard deviation of the true residuals; for the mean image and the first image of the set these are
        $5.92 \times 10^{-2}$ and $8.49 \times 10^{-2}$, respectively.}
        \label{fig:sat_iterm2_50}
\end{figure}

The second problem concerns edge effects and can be serious for images containing important details near their borders.
In this case the
circulant approximation used in model~(\ref{eq:modela}) is not appropriate and edge effects are to be expected. In particular,
the mean image $\zetah(m,n)$ of Eq.~(\ref{eq:solution2}) may have no relationship with the true image.
We have to stress that this kind of problem is unavoidable and is shared by any other method.
Its classical solution is the windowing of image $\gb$ to reduce edge discontinuities.

\citet{viob03} show that a remarkable reduction of the edge effects in deblurring operations, that
causes little distortion in the final result,
can be obtained if, instead of $\gb$, we use $\ob = \gb \, \cdot \hb$ (``$~\cdot~$'' denotes
element wise multiplication) and $\hb$ is a modification of
the classical Hanning window
\begin{equation} \label{eq:window}
h(m,n) = \left\{
\begin{array}{ll}
0.25 \times \alpha \times \beta, & 0 \le m,n < N_w; \\
0.25 \times \alpha \times \beta, & N-N_w \le m,n < N; \\
1 & {\rm otherwise}.
\end{array}
\right.
\end{equation}
The parameters $\alpha = [1-\cos(m \pi/N_w)]$ and $\beta = [1-\cos(n \pi/N_w)]$
are chosen so that
the pixels in the central subimage are not modified and the image has continuous first derivatives
at the edges (see Fig.~\ref{fig:windows}). This window approaches the classical
two-dimensional Hanning window as $N_w \rightarrow N/2$  and tends to the rectangular
window as $N_w \rightarrow 0$. The parameter $N_w$ thus determines the filtering
characteristics, in particular the frequency pass-band, of the window. Its ``optimal'' value depends on
many factors such as the noise level, the form of the PSF and the specific realization of the
process. For a Gaussian PSF, simulation results \citep{viob03} show that values of three or four times the dispersion of the PSF
provide reasonably good results. The only unavoidable consequence of the procedure is that
a border of thickness $N_w$ has to be removed from the final deblurred image.

According to these results, $\oh(m,n)$  should permit the use of~(\ref{eq:solution2}) also in
situations where the circulant approximation for the blurring operator is not appropriate. We test this
via numerical simulations.

\section{Some numerical experiments} \label{sec:experiments}

We present the results of some numerical simulations to show the flexibility and
reliability of the methodology presented above.
Figs.~\ref{fig:sat_iterc1_50}-\ref{fig:star_iterm2_50} show the deblurring results for
an extended object and a set of point-like objects. We have chosen these particular
examples since, being non-smooth functions, their restoration is a difficult problem.
Eight images are available for each object. In each case the PSF is a bidimensional Gaussian with
dispersion along the major axis set
to twelve pixels, and to four pixels along the minor axis. Their inclinations take equispaced
values in the range $0^{\circ} - 160^{\circ}$. Gaussian white noise is added with a standard deviation of
$2 \%$ of the maximum value of the blurred images. Two deblurring methods have been used: a classical {conjugate
gradient} method for LS problems (CGLS) \citep{bjo96}, and a modified residual norm steepest descent method (MRNSD)
that can be used to enforce nonnegativity of the solution \citep{NaSt00}.
For comparison, the deblurring of the first image of
the set (that with the PSF with major axis inclined at $0^{\circ}$) is also
shown. In every case the improvement due to the use of the eight images is evident.

\begin{figure}
        \resizebox{\hsize}{!}{\includegraphics{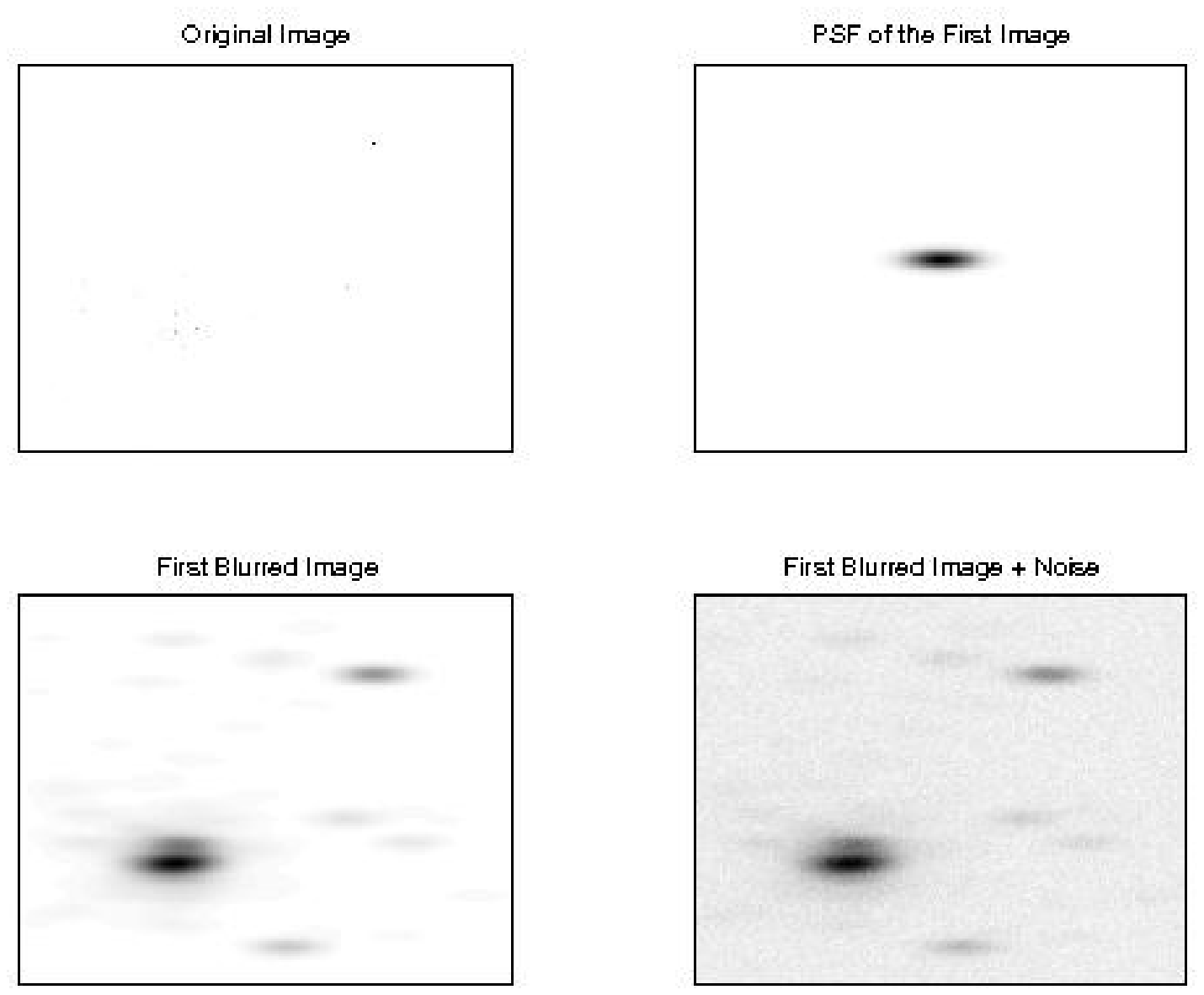}}
        \caption{Original image, blurred version and
        blurred + noise version of the first image of the set (see text) and corresponding PSF. The
        images are $256 \times 256$ pixels.}
        \label{fig:star_iterc1_50}
\end{figure}
\begin{figure}
        \resizebox{\hsize}{!}{\includegraphics{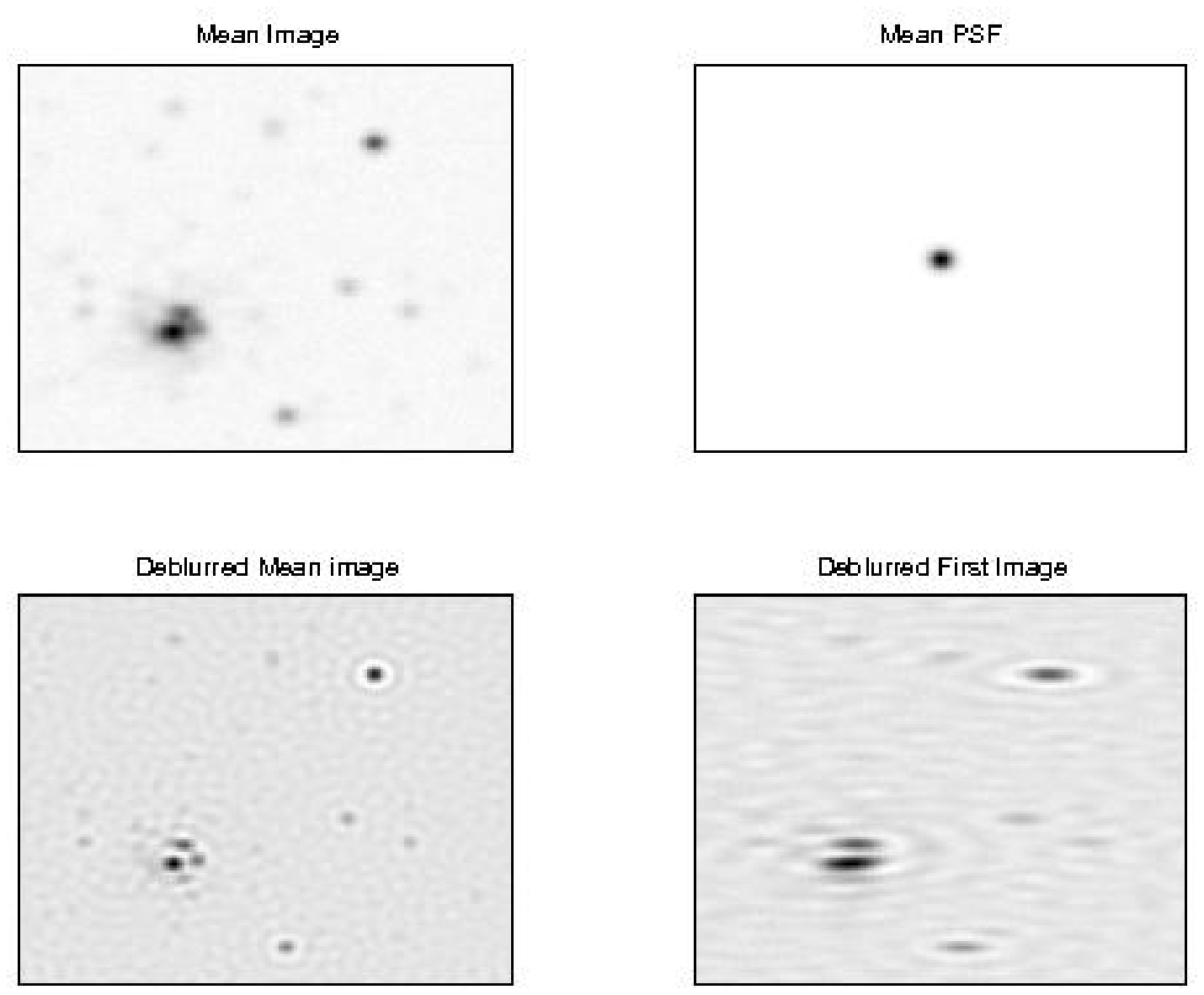}}
        \caption{Mean image $\zeta(m,n)$ and corresponding PSF for the image shown in
	  Fig.~\ref{fig:star_iterc1_50}, deblurred version of $\zeta(m,n)$,
        and deblurred version of the first image of the set (see text). The
        deblurring was done with CGLS. The images shown are the ones with the smallest
	  standard deviation of the true residuals; for the mean image and the first image of the set these are
        $6.09 \times 10^{-3}$ and $6.17 \times 10^{-3}$, respectively.}
        \label{fig:star_iterc2_50}
\end{figure}
\begin{figure}
        \resizebox{\hsize}{!}{\includegraphics{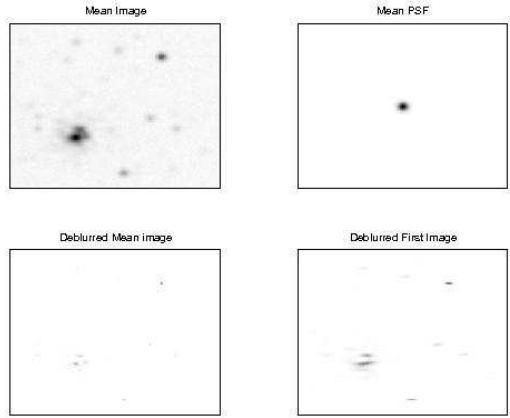}}
        \caption{Mean image $\zeta(m,n)$ and corresponding PSF for the image shown in
	  Fig.~\ref{fig:sat_iterc1_50}, deblurred version of $\zeta(m,n)$,
        and deblurred version of the first image of the set (see text). The
        deblurring was done with MRNSD ($1000$ iterations). The images shown are the ones with the smallest
	  standard deviation of the true residuals; for the mean image and the first image of the set these are
        $5.52 \times 10^{-3}$ and $6.02 \times 10^{-3}$, respectively.}
        \label{fig:star_iterm2_50}
\end{figure}

The two examples just presented do not suffer edge effects. The objects
are in the center of the images and their borders show only ``dark sky''. A more difficult example is provided by
images with details near their borders since in this case the circulant approximation for the blurring operator is incorrect.
In order to show the results obtainable in this case, we consider
observations of the Cosmic Microwave Background (CMB) \citep[e.g., see ][]{vioa03}. Although the PSFs considered
in this work are different from those typical of CMB observations, we have chosen this example
because the CMB signal is expected to be the realization of a stationary stochastic process covering the whole sky.
It represents one of the most unfavourable situations in the deblurring of astronomical images.
Figs.~\ref{fig:cmb1_2}-\ref{fig:cmb2_10}
show that, despite the difficulty of the problem, the application of the windowing procedure presented in
Sect.~\ref{sec:problems} can provide good results.
The experiments have been carried out under the same conditions as the previous ones, with the difference that
two different levels of noise have been tested ($2\%$ and $10\%$ of the dispersion of the blurred signal), and
the method used in the deblurring is the classical Tikhonov approach. The CMB maps have been simulated as described
in \citet{vioa03}. Again the results are reasonably good.

\begin{figure}
        \resizebox{\hsize}{!}{\includegraphics{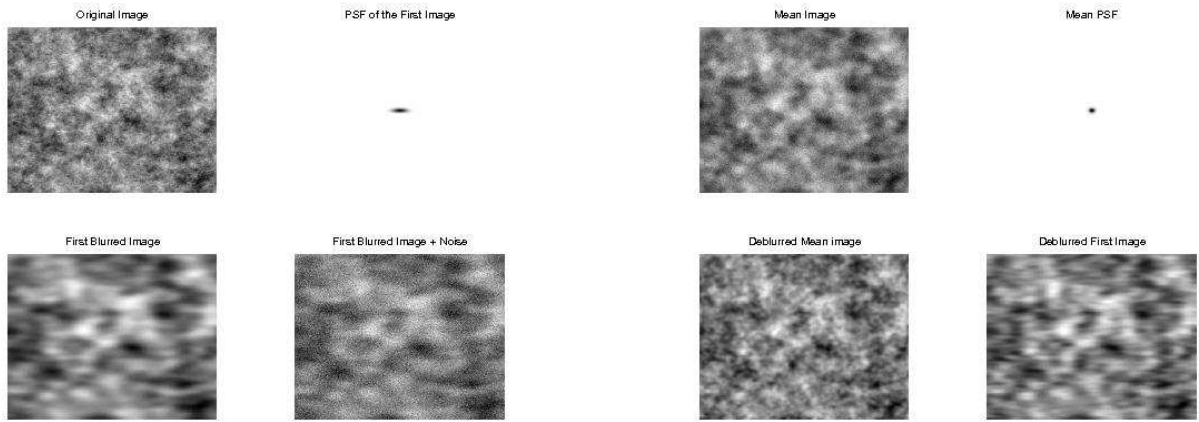}}
        \caption{Original image, blurred version and
        blurred + noise version of the first image of the set (see text) and corresponding PSF. The
        images are $340 \times 340$ pixels. Noise of S/N = 2.}
        \label{fig:cmb1_2}
\end{figure}
\begin{figure}
        \resizebox{\hsize}{!}{\includegraphics{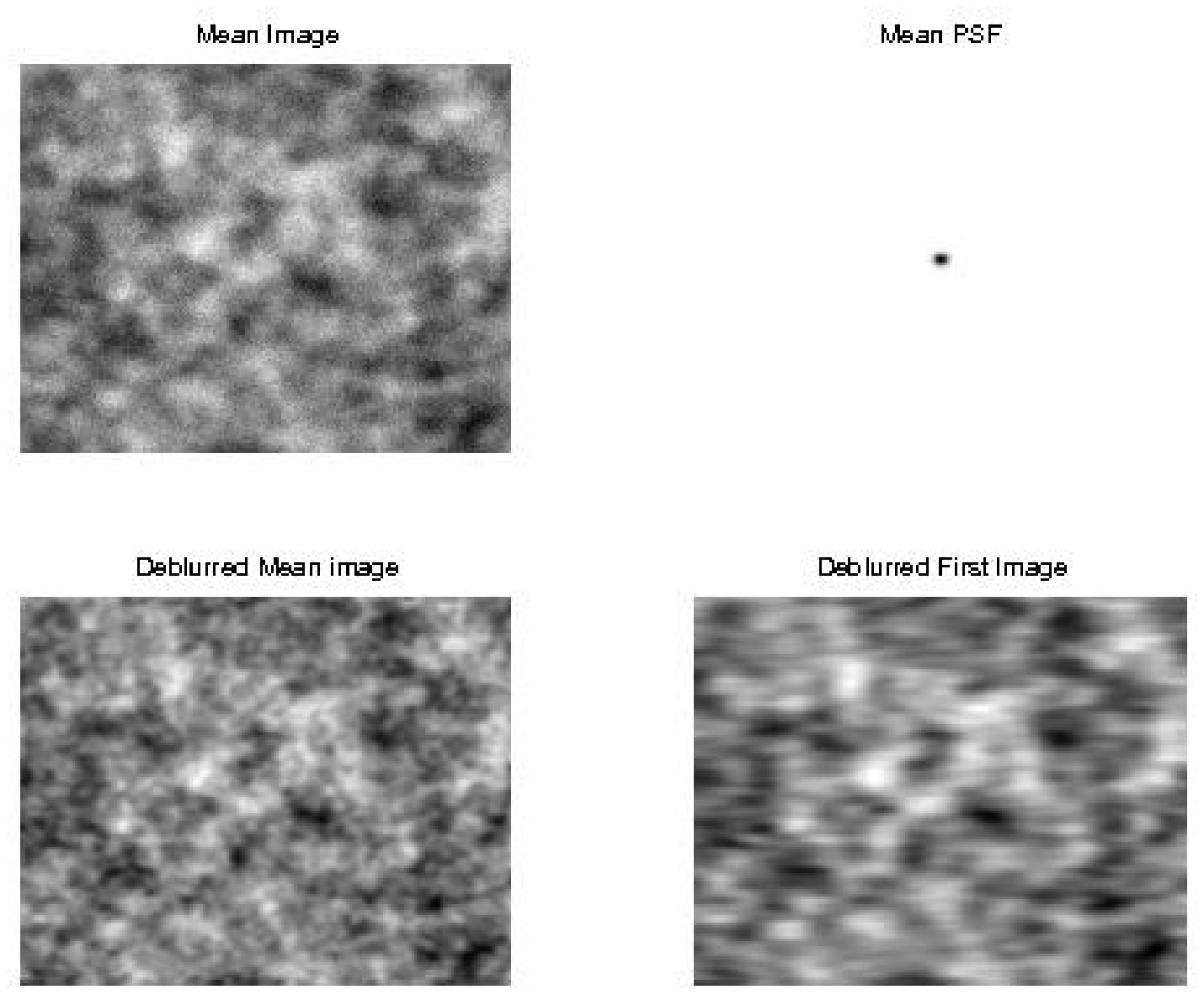}}
        \caption{Mean image $\zeta(m,n)$ and corresponding PSF for the image shown in
	  Fig.~\ref{fig:cmb1_2}, deblurred version of $\zeta(m,n)$,
        and deblurred version of the first image of the set (see text). The
        deblurring was done with the classical Tikhonov method and the windowing procedure explained in the text
	  ($N_w=30$ pixels). The images shown are the ones with the smallest
	  standard deviation of the true residuals that for the mean image and the first image of the set are
        $0.40$ and $0.52$, respectively.}
        \label{fig:cmb2_2}
\end{figure}
\begin{figure}
        \resizebox{\hsize}{!}{\includegraphics{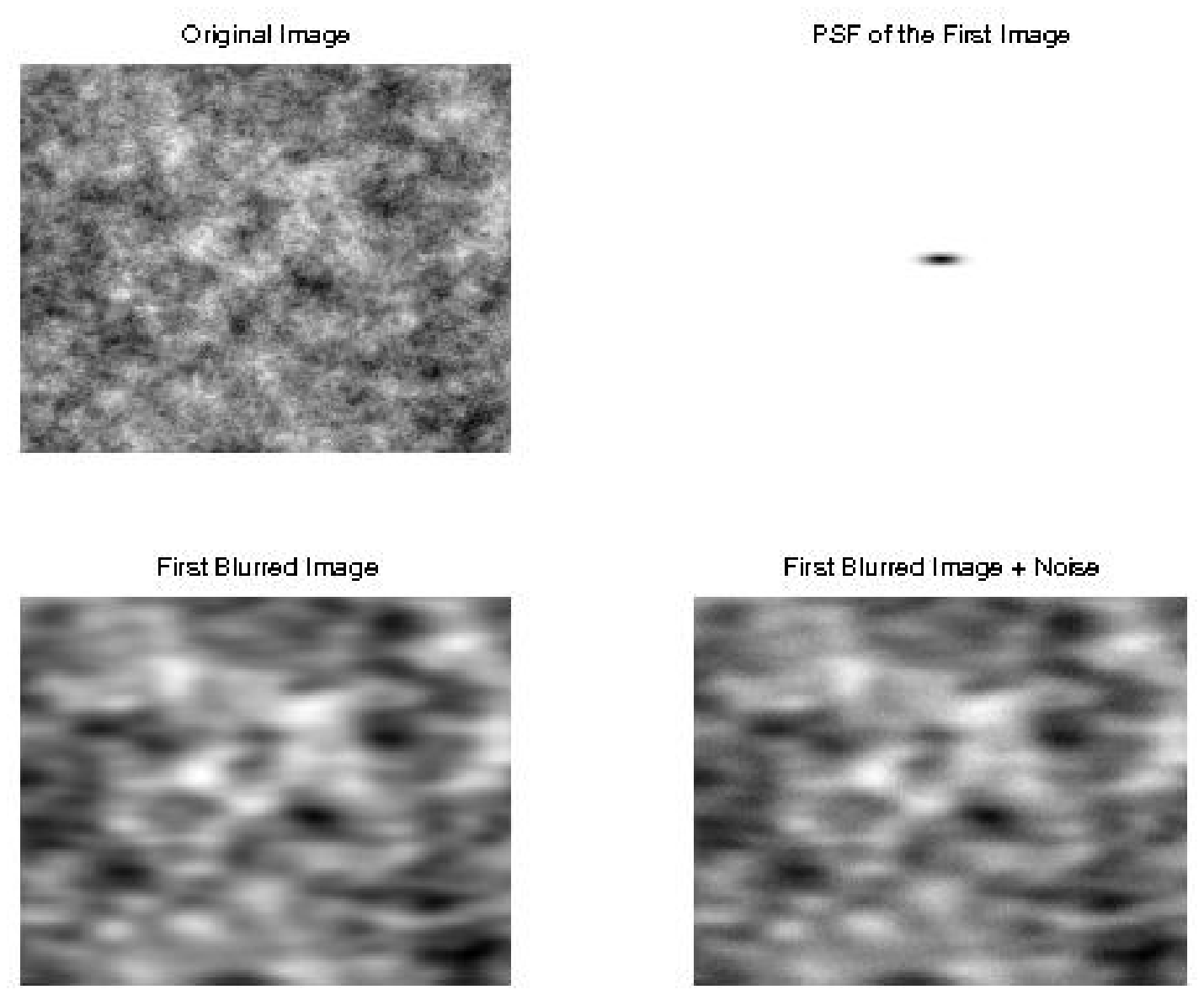}}
        \caption{Original image, blurred version and
        blurred + noise version of the first image of the set (see text) and corresponding PSF. The
        images are $340 \times 340$ pixels. Noise of S/N = 10.}
        \label{fig:cmb1_10}
\end{figure}
\begin{figure}
        \resizebox{\hsize}{!}{\includegraphics{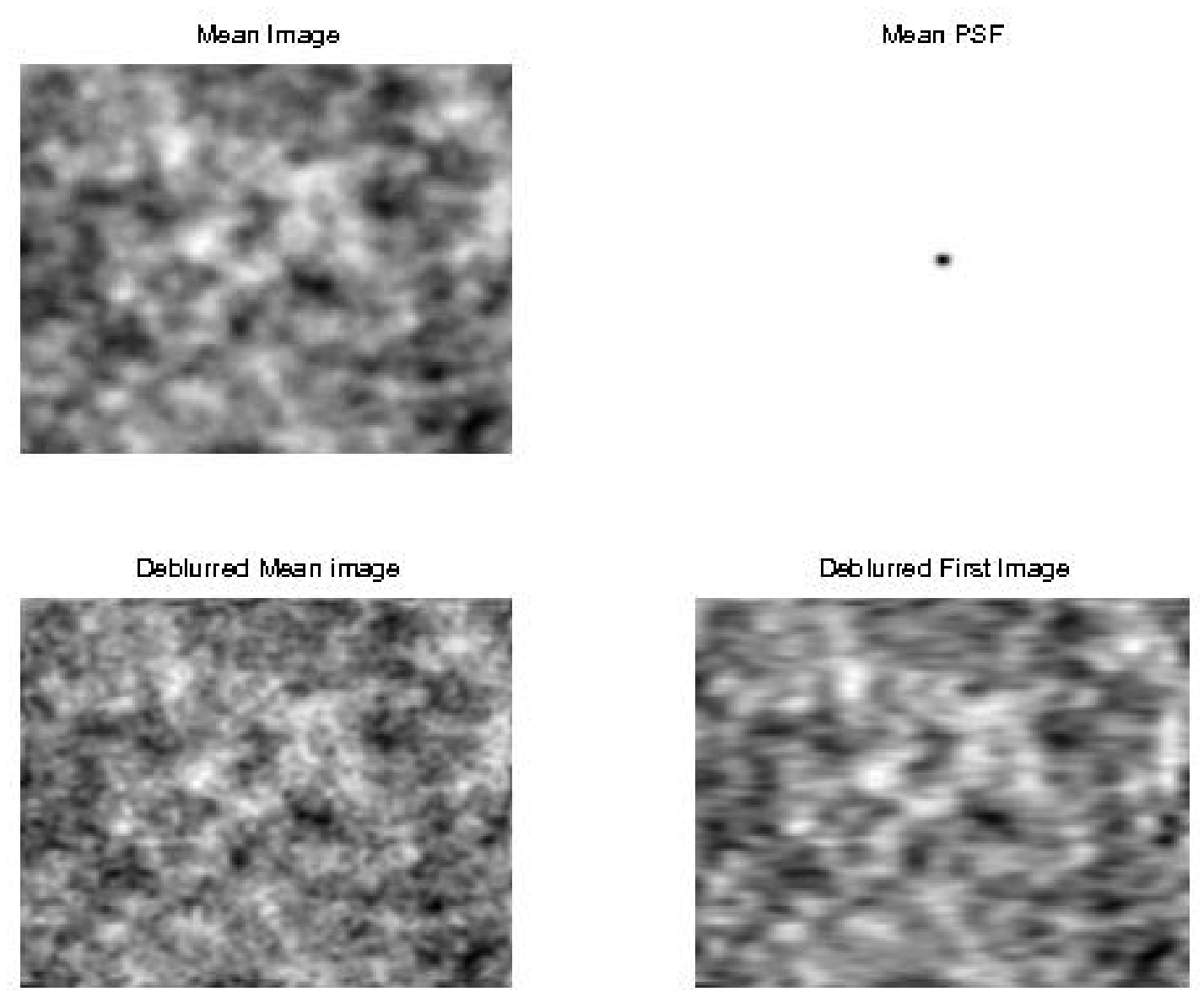}}
        \caption{Mean image $\zeta(m,n)$ and corresponding PSF for the image shown in
	  Fig.~\ref{fig:cmb1_10}, deblurred version of $\zeta(m,n)$,
        and deblurred version of the first image of the set (see text). The
        deblurring was done with the classical Tikhonov method and the windowing procedure explained in the text
	  ($N_w=30$ pixels). The images shown are the ones with the smallest
	  standard deviation of the true residuals that for the mean image and the first image of the set are
        $0.35$ and $0.48$, respectively.}
        \label{fig:cmb2_10}
\end{figure}

Our examples have shown that the use of $\zeta(m,n)$ coupled with standard deblurring algorithms
can remarkably improve the results. However, it is still necessary to check the effective gain in
computational cost. Figs.~\ref{fig:comp_sat50} and ~\ref{fig:comp_star50} show that the convergence rate of the MRNSD
algorithm applied to $\zeta(m,n)$ is faster than that of the PL or ISRA methods applied to $\vartheta(m,n)$
(the cost per iteration of each method is essentially the same).
We remark that MRNSD did not perform very well when applied to the more ill-conditioned
problem involving $\vartheta(m,n)$.  This is not too surprising since MRNSD is essentially
a steepest descent method, which is known to have poorer convergence properties for
more ill-conditioned problems.  It has been shown in \citet{NaSt00} that preconditioning
can dramatically increase the convergence rate for MRNSD, but it is a nontrivial process
to choose an appropriate preconditioner, especially for severely ill-conditioned problems;
see \citet{NaSt00} for further details.

Furthermore, to check
that this result is not due to the kind of algorithm used, Figs.~\ref{fig:comp_iter_sat50}, and
\ref{fig:comp_iter_star50} compare the convergence rate when the PL and ISRA methods are applied to both $\zeta(m,n)$
and $\vartheta(m,n)$. Again,
it is evident that the use $\zeta(m,n)$ can improve the convergence rate of the algorithms.

\begin{figure}
        \resizebox{\hsize}{!}{\includegraphics{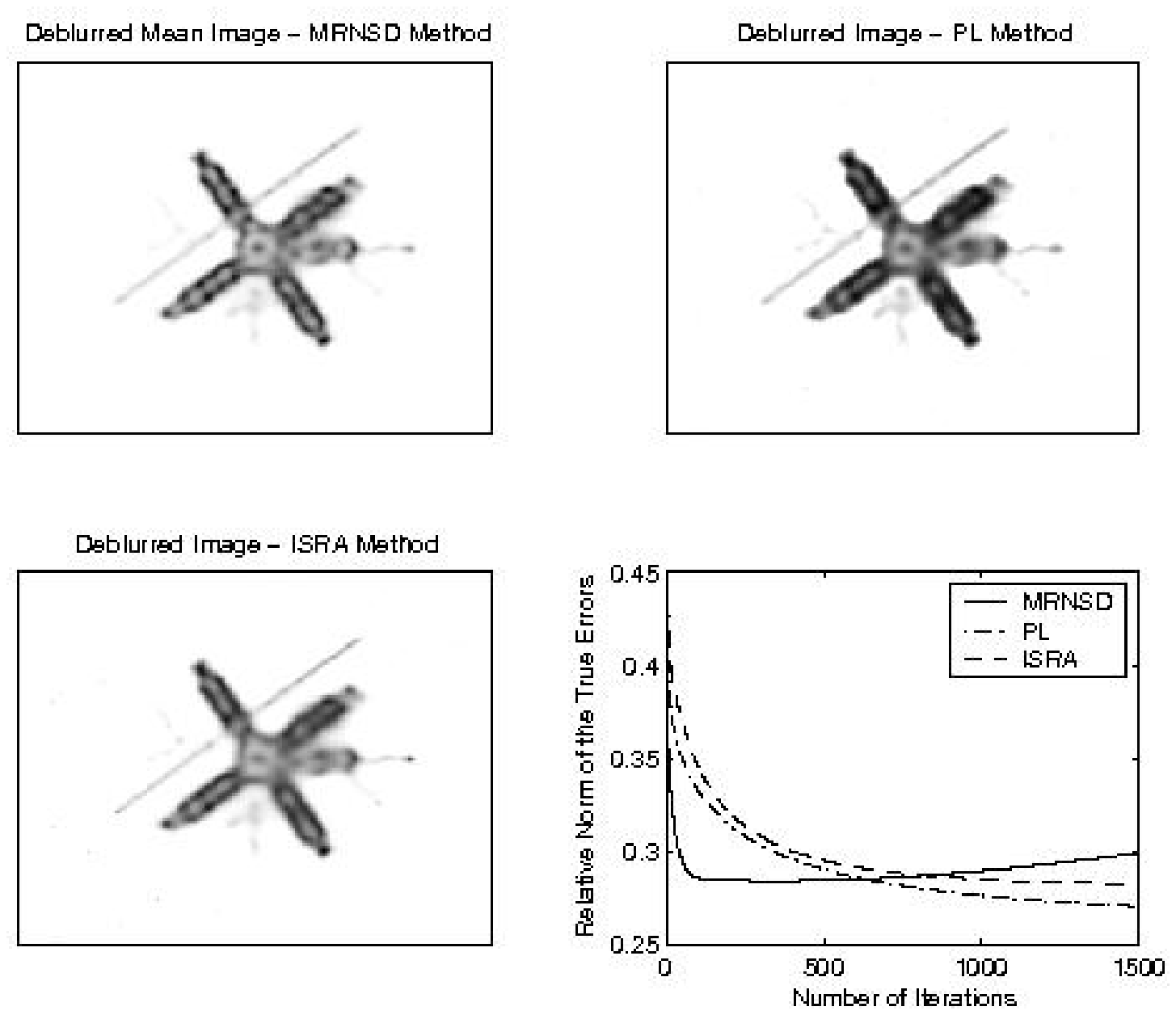}}
        \caption{Deblurred images, corresponding to the image shown in
	  Fig.~\ref{fig:sat_iterc1_50}, obtained through the MRNSD method applied to $\zeta(m,n)$ , and the PL, and ISRA
        methods applied to $\vartheta(m,n)$.
	  The images shown are the ones with the smallest
	  standard deviation of the true residuals that are
        $5.92 \times 10^{-2}$, $5.39 \times 10^{-2}$, and $5.63 \times 10^{-2}$ respectively.
	  The convergence curves for each method is also shown: MRSND converges after $346$ iterations, whereas
        both PL and ISRA do not converge after $1500$ iterations.}
        \label{fig:comp_sat50}
\end{figure}
\begin{figure}
        \resizebox{\hsize}{!}{\includegraphics{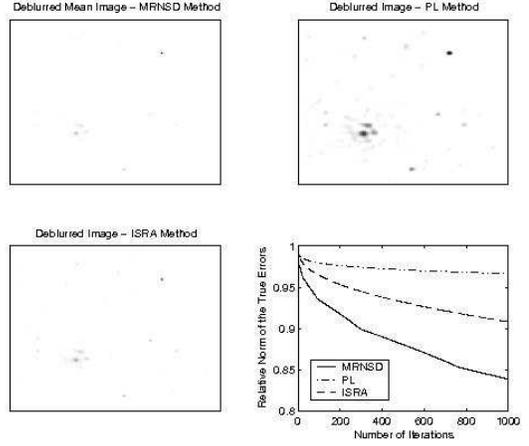}}
        \caption{Deblurred images, corresponding to the image shown in
	  Fig.~\ref{fig:star_iterc1_50} obtained through the MRNSD method applied to $\zeta(m,n)$ , and the PL, and
        ISRA methods applied to $\vartheta(m,n)$.
	  The images shown are the ones with the smallest
	  standard deviation of the true residuals that are
        $5.52 \times 10^{-3}$, $6.05 \times 10^{-3}$, and $5.69 \times 10^{-2}$ respectively.
	  The convergence rate for each method is also shown: none of the methods converged before 1000
        iterations.}
        \label{fig:comp_star50}
\end{figure}
\begin{figure}
        \resizebox{\hsize}{!}{\includegraphics{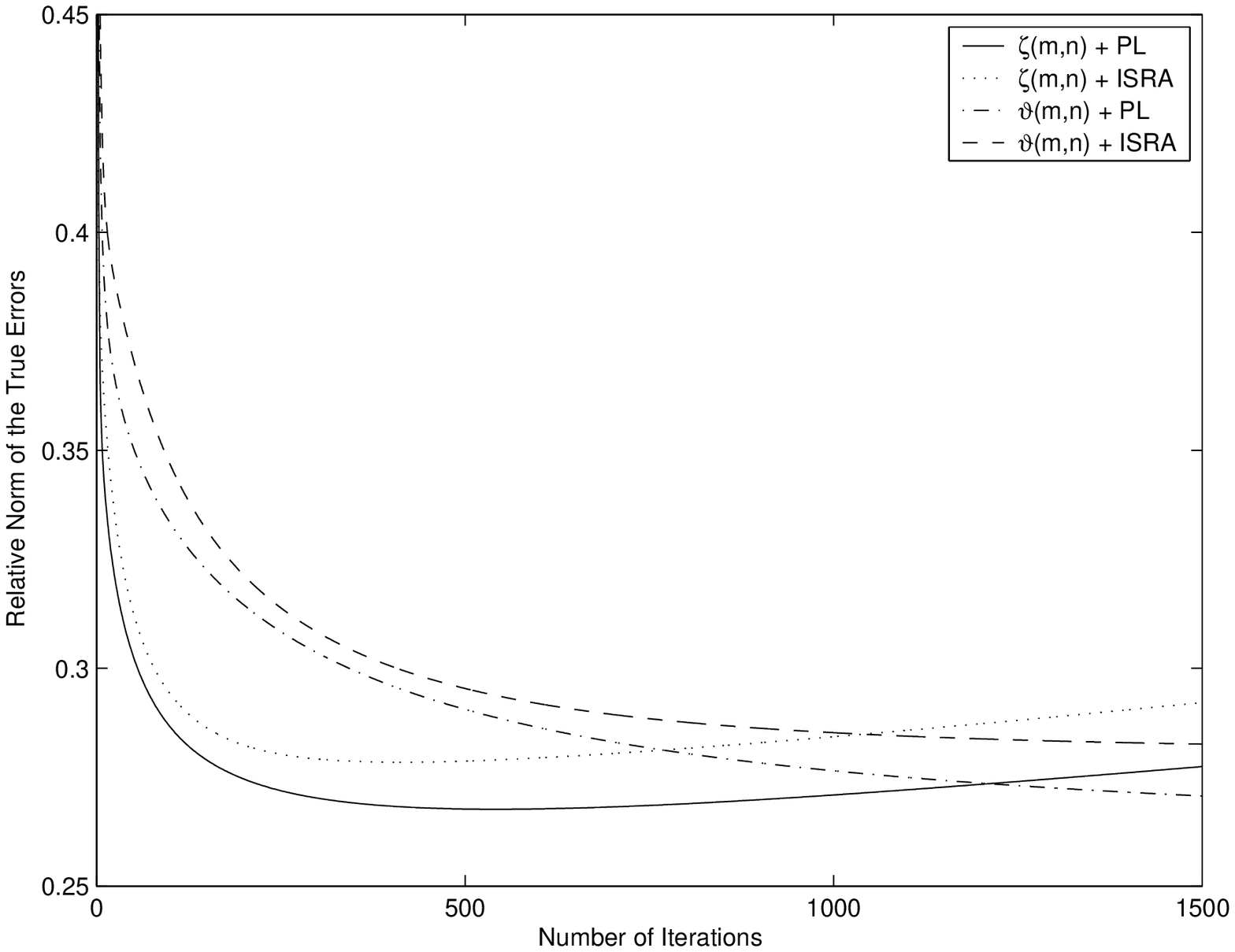}}
        \caption{Convergence rate, corresponding to the image shown in
	  Fig.~\ref{fig:sat_iterc1_50}, for the PS and ISRA methods applied to both $\zeta(m,n)$ and $\vartheta(m,n)$.}
        \label{fig:comp_iter_sat50}
\end{figure}
\begin{figure}
        \resizebox{\hsize}{!}{\includegraphics{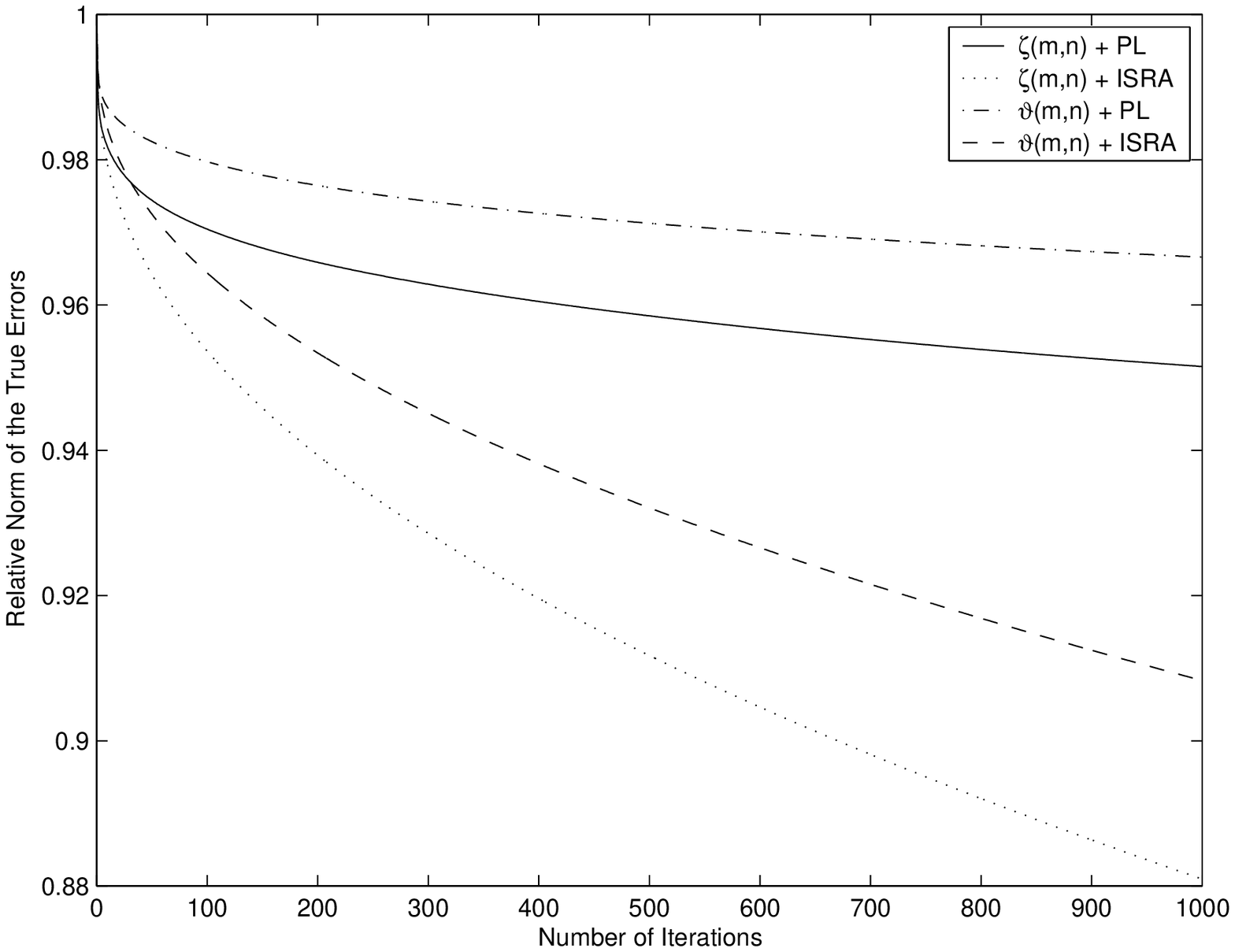}}
        \caption{Convergence rate, corresponding to the image shown in
	  Fig.~\ref{fig:star_iterc1_50}, for the PS and ISRA methods applied to both $\zeta(m,n)$ and $\vartheta(m,n)$.}
        \label{fig:comp_iter_star50}
\end{figure}

\section{Extension to images with different noise levels}

So far we have assumed that the noise level is the same in all the images ($\sigma_{w_j} = \sigma_w$),
a condition that allows the combination of different images
to provide the most interesting results. Moreover, this requirement is often satisfied in practice.
For example, with LBT the images taken at different orientations of the telescope are expected to be
characterized by the same noise level.

The extension of the method to images with different noise levels is straightforward.
If the different $\sigma_{w_i}$ are known, one may account for the different variabilities by
changing equation~(\ref{eq:min2}) to
\begin{equation}
\fb = \arg\min\, \sum_{j=1}^{p} \Vert\, (\,\Ab_j \fb - \gb_j\,)/\sigma_{w_j} \,\Vert^2 ,
\end{equation}
or, alternatively,
\begin{equation}
\fb = \arg\min \,\sum_{j=1}^{p} \Vert\, \amatb_j \fb - \varrhob_j \,\Vert^2 ,
\end{equation}
where $\amatb_j$ and $\varrhob_j$ represent, respectively, the matrix $\Ab_j$ and array $\gb_j$ with their
elements divided by $\sigma_{w_j}$. The rest of the procedure remains as described in
Sect.~\ref{sec:efficient}. For example, Eq.~(\ref{eq:solution2k}) becomes
\begin{equation}
\kmath_{j_0}(m,n) = \max \,[ \,\kmath_1(m,n), \kmath_2(m,n), \ldots, \kmath_p(m,n)\, ],
\end{equation}
where $\kmath_j(m,n) = \Kh_j(m,n)/\sigma_{w_j}$.

\section{Concluding remarks} \label{sec:conclusions}

We have considered the problem of simultaneously deblurring a set of images of a fixed object blurred by
different space-invariant PSFs. Although currently available methods
seem to provide good results, they are comptutationally expensive.
We have developed a method, based on a LS approach, to efficiently
transform a multi-image deblurring problem into a single-image one. This approach
provides substantial savings in computational requirements and can be implemented using standard
currently available numerical algorithms. These conclusions are confirmed by our numerical experiments.

But despite these encouraging results, some questions are still open. In particular,
regularization parameter selection methods (such as GCV for the Tikhonov approach)
have to be extended to deal with the correlated noise component of the mean image.
This is especially important for the deblurring of random fields where the statistical properties
of the field are more important than a particular realization.

Another question is the generalization of the proposed method to non-Gaussian noise.
For example, for Poisson noise good results have been obtained
with the {maximum-likelihood} estimate
\begin{align}
\fb &= \arg\max \,\sum_{j=1}^p \sum_{m,n=0}^{N-1} \nonumber \\
 & [ \,g_j(m,n) \ln (\Ab_j \fb)(m,n) - (\Ab_j \, \fb)(m,n) ] ,
\end{align}
\citep{cor00, ber00a}.

\begin{acknowledgements}
We thank Prof. M. Bertero (Universit\'a di Genova) for useful discussions.
\end{acknowledgements}

\appendix
\section{Multi-Frame Deblurring and Weighted Least-Squares} \label{sec:ls}

In this appendix the multi-frame
deblurring problem presented in the previous sections is shown to be related to
(column) weighted LS.

In the multi-frame deblurring problem, we consider the
LS problem (we omit regularization, but this
can be easily incorporated into the discussion):
\begin{equation}
   \min\left\| \left[ \begin{array}{c}
                        \Ab_1 \\ \Ab_2 \\ \vdots \\ \Ab_p
                      \end{array}
               \right] \fb -
               \left[ \begin{array}{c}
                        \gb_1 \\ \gb_2 \\ \vdots \\ \gb_p
                      \end{array}
               \right] \right \|_2
\end{equation}

Assuming each $\Ab_j$ is block circulant, and can be diagonalized
by the unitary Fourier transform matrix, then this problem
can be transformed to the equivalent LS problem:
\begin{equation}
   \min\left\| \left[ \begin{array}{c}
                        \Lambdab_1 \\ \Lambdab_2 \\ \vdots \\ \Lambdab_p
                      \end{array}
               \right] \fbh -
               \left[ \begin{array}{c}
                        \gbh_1 \\ \gbh_2 \\ \vdots \\ \gbh_p
                      \end{array}
               \right] \right \|_2
\end{equation}
where $\Lambdab_j = \mbox{diag}(\lambda_1^{(j)}, \lambda_2^{(j)},
\ldots, \lambda_N^{(j)})$ is a diagonal matrix containing the eigenvalues
of the matrix $\Ab_j$ (i.e., this is just the DFT of the PSF,
or in the notation of the main text $\Kbh_j$).

Now, following the arguments in Sect.~\ref{sec:efficient}, let
$\displaystyle
  \delta_i = \max_{1 \leq j \leq p}\{\lambda_i^{(j)}\}\,,
$
$i = 1, 2, \ldots, N$.  Define the diagonal matrix:
\begin{equation}
  \Deltab = \mbox{diag}(\delta_1, \delta_2, \ldots, \delta_N)\,,
\end{equation}
and consider the (column) weighted LS problem:
\begin{equation}
   \min\left\| \left[ \begin{array}{c}
                        \Lambdab_1 \Deltab^{-1} \\
                        \Lambdab_2 \Deltab^{-1} \\ \vdots \\
                        \Lambdab_p \Deltab^{-1}
                      \end{array}
               \right] \Deltab \fbh -
               \left[ \begin{array}{c}
                        \gbh_1 \\ \gbh_2 \\ \vdots \\ \gbh_p
                      \end{array}
               \right] \right \|_2
\end{equation}
We will write this LS problem as:
\begin{equation}
   \min || \,\Db \Deltab \fbh - \gbh\,||_2\,,
\end{equation}
where
\begin{equation}
   \Db = \left[ \begin{array}{c}
                        \Lambdab_1 \Deltab^{-1} \\
                        \Lambdab_2 \Deltab^{-1} \\ \vdots \\
                        \Lambdab_p \Deltab^{-1}
                      \end{array}
               \right]\,.
\end{equation}
The normal equations version of this LS problem is:
\begin{equation}
   \Deltab^* \Db^*\Db \Delta \fbh = \Deltab^* \Db^* \gbh\,,
\end{equation}
or, equivalently,
\begin{equation}
\label{NE}
   \Db^*\Db \Deltab \fbh = \Db^* \gbh\,,
\end{equation}
where $\Db^*$ is the complex conjugate transpose of $\Db$.

Now, observe that $\Db^*\Db$ is perfectly well conditioned, and thus
can be easily inverted.  To see this, note that for all $i$,
there exists $j_0 = j_0(i)$ such that
\begin{equation}
   \delta_i = \lambda_i^{(j_0)}\,, \quad \mbox{and} \quad
   |\,\lambda_i^{(j_0)}\,| \geq |\,\lambda_i^{(j)}\,|\,, \; j \neq j_0\,.
\end{equation}
The $i$th diagonal entry of $\Db^*\Db$ is
\begin{equation}
  \sum_{j=1}^p \frac{|\,\lambda_i^{(j)}\,|^2}{|\,\delta_i\,|^2} =
  1 + \sum_{j \neq j_0} \frac{|\,\lambda_i^{(j)}\,|^2}{|\,\lambda_i^{(j_0)}\,|^2}
\end{equation}
and thus each diagonal entry of $\Db^*\Db$ can be bounded by
\begin{equation}
  1 \leq \sum_{j=1}^p \frac{|\,\lambda_i^{(j)}\,|^2}{|\,\delta_i\,|^2} \leq p\,.
\end{equation}
Thus $\Db^*\Db$ is perfectly well conditioned.

Since $\Db^*\Db$ is perfectly well conditioned, there is no danger
in transforming the normal equations given in equation (\ref{NE}) to
\begin{equation}
   \Deltab \fbh = (\Db^*\Db)^{-1}\Db^* \gbh\,,
\end{equation}
which is equivalent to Eq.~(\ref{eq:solution2}).

A vast literature is available on column weighted LS;
see, for example, \citet{gol96}, \citet{law95}, and \citet{bjo96}.

\end{document}